\pdfoutput=1
\documentclass[aps,prl,twocolumn,showpacs,longbibliography,floatfix,superscriptaddress]{revtex4-1}
\usepackage{graphicx}
\usepackage{bm}
\usepackage{amsmath}
\usepackage{amssymb}
\usepackage{euscript}
\usepackage{verbatim}
\usepackage{setspace}
\usepackage{xcolor}
\usepackage{amsfonts}
\usepackage{braket}
\usepackage{subfigure}
\usepackage[normalem]{ulem}

\newcommand{\be}{\begin{equation}}
\newcommand{\ee}{\end{equation}}
\newcommand{\bea}{\begin{eqnarray}}
\newcommand{\eea}{\end{eqnarray}}

\allowdisplaybreaks

\newcommand{\ES}[1]{{\color{red}#1}}

\begin{document}

	\title{
 Identifying an environment-induced localization transition from entropy and conductance
	}
	
	\author{Zhanyu Ma}
	\affiliation{School of Physics and Astronomy, Tel Aviv University, Tel Aviv 6997801, Israel}

	\author{Cheolhee Han}
	\affiliation{School of Physics and Astronomy, Tel Aviv University, Tel Aviv 6997801, Israel}
	
	\author{Yigal Meir}
\affiliation{Department of Physics, Ben-Gurion University of the Negev, Beer-Sheva, 84105 Israel}
		\author{Eran Sela}
	\affiliation{School of Physics and Astronomy, Tel Aviv University, Tel Aviv 6997801, Israel}
	\begin{abstract}
Environment-induced localization transitions (LT) occur when a small quantum system 
interacts with a bath of harmonic oscillators. At equilibrium, LTs are accompanied by an entropy change,  signaling the loss of coherence.
 Despite extensive efforts, equilibrium LTs have yet to be observed.  Here, we demonstrate that ongoing experiments on double quantum dots that measure entropy using a nearby quantum point contact (QPC) realize the  celebrated spin-boson model and allow to measure the entropy change of its LT.  We find a Kosterlitz-Thouless flow diagram, leading to a universal jump in the spin-bath 
 interaction, reflected in a discontinuity in the zero temperature QPC conductance.
	\end{abstract}
	
	\maketitle

\title{}

\maketitle

\emph{Introduction---} 
Environment-induced localization transitions (LTs) occur when a small quantum system switches from coherent to incoherent dynamics due to its interaction with an infinite number of environmental degrees of freedom. A simple example of this is the spin-boson model~\cite{leggett1987dynamics,weiss2012quantum}. It was proposed already 40 years ago~\cite{chakravarty1982quantum,bray1982influence,schmid1983diffusion} 
that  when the coupling of the spin or two-level system to
the bath exceeds a certain threshold, the tunneling between the two levels vanishes. Despite numerous proposals to observe this phase transition in various mesoscopic~\cite{le2005unification, li2005hidden,tong2006signatures,goldstein2010population} and atomic~\cite{recati2005atomic,porras2008mesoscopic}
systems, or by tracking the dynamics of the quantum system~\cite{porras2008mesoscopic}, this LT has not been observed to date without external driving~\cite{magazzu2018probing}. This is largely due to the experimental difficulty of continuously tuning the coupling or altering the power-law spectrum of the bosonic bath. Different than dissipative phase transitions~\cite{kessler2012dissipative,fitzpatrick2017observation,casteels2017critical,minganti2018spectral,fink2018signatures,hwang2018dissipative} that occur out of equilibrium, LTs can be identified in their thermodynamic properties. Here, utilizing the fact that the entropy displays a characteristic change 
across the transition~\cite{bulla2005numerical} we demonstrate that  already existing experimental setups measuring the entropy of quantum dot (QD) systems~\cite{hartman2018direct,child2022entropy,child2022robust} can be employed to observe the hitherto elusive LT for the spin-boson model at equilibrium. 

\begin{figure}[]
	\includegraphics[width=1\columnwidth]{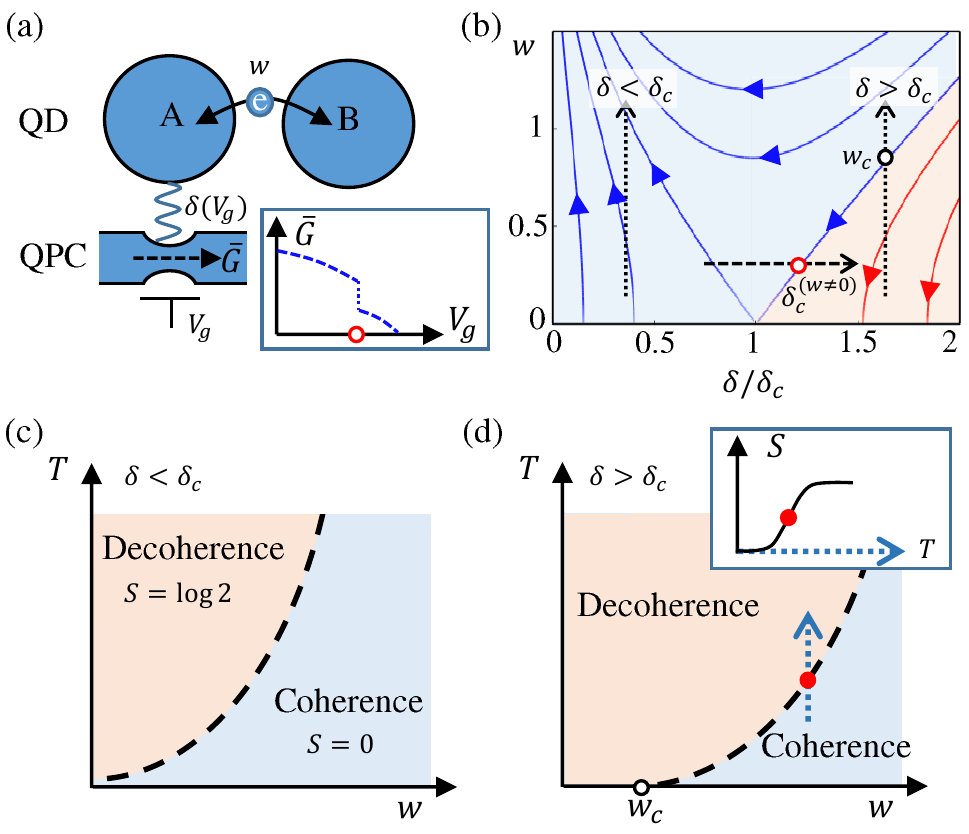}
	\caption{  (a) 
 Model: quantum dot (QD) $A$  tunnel coupled (via $w$) to QD $B$, here another QD, and 
 electrostatically coupled (as parametrized by $\delta$) to a quantum point contact (QPC). $\delta$ is tuned by a gate voltage $V_g$ and $\bar{G}$ is the average QPC conductance. The inset shows  a conductance jump obtained by changing $V_g$($\delta$). 
 (b) RG flow diagram. 
 In the red (blue) shaded area, $\delta$ flows to $\delta\ne 0$ ($\delta =0$). Two black dotted lines correspond to the $w$ axis of (c) and (d) and the black dashed line corresponds to the $V_g$ axis of the inset of (a).  (c) At $\delta<\delta_c$, there is no quantum phase transition tuned by $w$. (d) For $\delta>\delta_c$ at $T=0$, there appears a LT at $w=w_c$ (empty circle), characterized by an entropy jump.
%
 }\label{fig:1}
\end{figure}

Our proposed  realization of the two-level system is a double dot (DD)  containing a single electron, having a pair of states denoted $\{ |01\rangle, |10\rangle\}$, with the electron being in the right or  left QD, respectively (the spin of the electron is irrelevant).
The role of the bath is played by a nearby quantum point contact (QPC), whose transmission is controlled by gate voltage $V_g$, which 
acts
as a charge detector~\cite{field1993measurements}
of QD $A$, see Fig.~\ref{fig:1}(a). 
Below, we relate the QPC-QD electrostatic coupling to an effective change of the scattering phase shift $\delta$ in the QPC, occurring as an electron enters QD $A$.

The decoupled system ($\delta=0$) undergoes a $k_B \log 2$ entropy drop (where $k_B$ is the Boltzmann constant, set to unity in the following) as the temperature is lowered below the DD
tunneling amplitude $w$. Indeed, the symmetric 
DD transitions from the high-temperature state, described by the diagonal density matrix  
$(|01\rangle \langle 01 |+|10\rangle \langle 10 | )/2$, to the coherent ground state $(|01\rangle+|10\rangle)/\sqrt{2}$. We predict that for a nonzero coupling to the bath, $\delta>0$, the temperature scale for this incoherent-coherent crossover decreases, and eventually vanishes at the LT at some $\delta = \delta_c$. This can be understood  from the Anderson  orthogonality catastrophe~\cite{anderson1967infrared,aleiner1997dephasing, goldstein2010population}. For $\delta > \delta_c$ 
the orthogonality between the many-body wave functions of the QPC for the QD states $|01\rangle$ and $|10\rangle$ effectively turns the tunneling amplitude $w$ 
to zero, see Fig.~\ref{fig:1}(b,d), reminiscent of the Zeno effect~\cite{misra1977zeno,hackenbroich1998quantum}, and the entropy remains $\log 2$ down to zero temperature. 
 We find that both $w$ and the effective electrostatic interaction $\delta$ are scaling variables of the LT described by the renormalization group (RG) flow diagram of a Kosterlitz-Thouless (KT) transition, see Fig.~\ref{fig:1}(b).
While it may be difficult to tune $\delta$, the LT can be driven as a function of tunneling coupling $w$, which can be readily tuned by a gate. As seen in Fig.~\ref{fig:1}(d) there is 
a critical tunneling amplitude $w_c$ below which the entropy remains finite at zero temperature, while coherence develops for $w>w_c$, manifested as a drop of the entropy to zero at low temperature.

From this flow diagram, we can see that the effective QPC-QD interaction $\delta_{\rm{{eff}}}$ at low temperature, being the destination of the flow diagram, changes discontinuously depending on its bare value, between a finite value $\delta_{\rm{{eff}}} \geq \delta_c$ in the incoherent phase, and $\delta_{\rm{{eff}}}=0$ in the coherent phase. This universal step is the analog of the discontinuity in the superfluid density in the standard KT transition due to vortices
. In our system, unexpectedly, 
this is reflected as a sudden change of the QPC conductance as $T\to 0$, see inset of Fig.~\ref{fig:1}(a).

\emph{
Model---} As depicted in Fig.~\ref{fig:1}(a), we consider a DD  electrostatically coupled to a QPC with Hamiltonian $H=H_{DD}+H_{QPC}
$. 
 $H_{DD}$, defined explicitly in Eq.~(\ref{eq:syslattuceDD}) below,
 describes two subsystems. Subsystem $A$ is a QD in the Coulomb blockade regime which accommodates only two charge states labeled by ${N}_A=0,1$, while subsystem $B$ could be, in principle, arbitrary. For simplicity, we  
 consider here the case when the subsystem $B$ is another QD (for more examples, see~[\onlinecite{SM}]). The two subsystems are connected via a tunneling 
 amplitude $w$.
 We use the Pauli matrix $\sigma^z = |1\rangle \langle 1 | -|0 \rangle \langle 0 | $ to denote the charge operator of QD A, $\hat{N}_A=(1+\sigma^z)/2$. For the symmetric DD system, $H_{DD}=w \sigma^x$.
 
  The QPC consists of a quantum wire running along the $x$ direction, 
  interrupted by a 
  potential barrier $V_{N_A}(x,y)$, 
  \bea
H_{QPC}&=&\int dx dy \sum_{s=\uparrow , \downarrow}  \Psi_s^\dagger(x,y) \bigg[-\frac{\hbar^2}{2m} \nabla^2 \nonumber\\ &+& V_{0}(x,y)  |0\rangle \langle 0 |+V_{1}(x,y)  |1\rangle \langle 1 |\bigg]\Psi_s(x,y).
\label{eq:QPC}
\eea
  As a consequence, the potential in the QPC suddenly switches between $V_{1}(x,y)$ and $V_{0}(x,y)$ as an electron tunnels in and out of QD $A$. An explicit model for $H_{QPC}$ is considered below in Eq.~(\ref
{eq:V}) where $V_{N_A}(x,y)=V(y)+V_{N_A}(x)$, with $V_{N_A}(x)=V_{N_A}(-x)$. In this case, 
for each transverse mode $n=0,1,2,\dots$ of the potential $V(y)$ the scattering matrix is diagonal and encoded by the even and odd phase shifts $\delta^{(e,n)}_{N_A}$, $\delta^{(o,n)}_{N_A}$. For more general cases see~\cite{SM}. For convenience, we label  parity, channel, and spin index collectively by a single index $i=\{e/o,n,s \}=1,\dots, i_{max}$, and also define the difference and average phase shifts in each channel $\delta_i = (\delta_0^{i}-\delta_1^{i})/2$ and $\bar{\delta}_i = (\delta_0^{i}+\delta_1^{i})/2$.



\emph{Mapping to the spin-boson model---} 
At energies close to the Fermi energy, the fermion fields $\psi_{n,s}(x)$ have left and right components~\cite{affleck1994fermi,gogolin2004bosonization}, $\psi_{n,s}(x)=e^{i k_{F,n} x} \psi_{R,n,s}+e^{-i k_{F,n} x} \psi_{L,n,s}$. It allows to define even and odd chiral fields $\psi_{e/o,n,s}=(\psi_{R,n,s}(x) \pm \psi_{{L,n,s}}(-x))/\sqrt{2}$ which we bosonize~\cite{kane2005lectures} into 
\bea
    \label{eq:O}
H_{QPC} =v_F \sum_{i} \left( 
\int \frac{dx}{4 \pi} (\partial_x \phi_i)^2-
\frac{\delta_i }{\pi}
\sigma^z  
\partial_x \phi_i(0) \right)+H_{ps}, 
\eea
with $[\phi_j(x), \partial_y \phi_k(y)]=-2\pi i \delta_{jk}\delta(x-y)$~\cite{kane2005lectures}.   
The second 
term $\propto \sigma_z$ describes the $N_A$-dependent potential. The last term $H_{ps}=-\frac{v_F}{\pi}  \sum_i \bar{\delta}_i 
\partial_x \phi_i(0)$ is a constant potential  
that can be removed by a unitary transformation $H \to U H U^\dagger$ with $U=e^{-i\sum_j \bar{\delta}_j  \phi_j(0)/\pi}$. 
We define 
\be
\label{eq:deltaeff}
\delta=\sqrt{ \sum_{i    }  \delta_i ^2},
\ee
and $\phi(x)=\frac{1}{\delta}\sum_i \delta_i \phi_i(x) $, along with $i_{max}-1$ orthogonal combinations $\{ \phi'_{i}, i=2,\dots,i_{max} \}$ which do not interact with $\sigma^z$. We obtain $H=H_{{\rm{eff}}}(\phi) + \sum_{i=2}^{i_{max}} H[\phi']$ where the LT is captured by the effective model
\bea
    \label{eq:O}
H_{{\rm{eff}}}(\phi) =  \frac{v_F}{4\pi} \int dx (\partial_x \phi)^2-\frac{v_F}{\pi}  
\delta \sigma^z  
\partial_x \phi(0)+w \sigma^x. 
\eea
This model is equivalent to the spin-boson model with an Ohmic bath~\cite{SM}. The term  $\propto \delta \sigma^z$ describes the interaction between the spin and the bosonic environment $\phi(x)$.


\emph{Anderson orthogonality catastrophe and LT---} 
One can apply a similar transformation $U'=e^{i \sigma^z \delta \phi(0)/\pi}$ to remove the interaction $\propto \delta \sigma^z$ from Eq.~(\ref{eq:O}). Then the tunneling term $\propto w \sigma^+ + h.c.$ gets `dressed' by a bath operator known as a boundary condition changing operator with scaling dimension $x_b = 2\left(\delta/\pi \right)^2$~\cite{affleck1994fermi,goldstein2010population}.
It reflects the orthogonality of the many-body ground states of the QPC for 
$N_A=0,1$. Thus the tunneling $w$ satisfies the RG equation $dw/dl=w(1-x_{tun}-x_b)$ where $x_{tun}=0$ is the bare scaling dimension of $w$. 
More generally, we find~\cite{SM}
\be
\label{eq:dwdl}
\begin{split}
\frac{dw}{dl}=&w\left(1-\left(\frac{\delta}{\delta_c}\right)^2\right),\quad \frac{d\delta}{dl}=-2\delta w^2,
\end{split}
\ee 
where $\delta_c=\pi/\sqrt{2}$ for the DD.
For small enough $w$, we see that  $w$ switches, upon increasing $\delta$, from being relevant to irrelevant at 
$\delta = \delta_c$. This critical interaction separates the strong interaction phase $\delta>\delta_c$ in which the coherent tunneling is  suppressed as in the Zeno effect, from the weak interaction phase $\delta<\delta_c$ with coherent tunneling. Equivalently, there is an energy scale that vanishes at the quantum phase transition~\cite{bulla2008numerical,SM} $T^*  \approx    w^{\pi^2/4 \delta_c (\delta_c - \delta)}$. More generally, also $\delta$ flows according to the celebrated KT flow diagram in Fig.~\ref{fig:1}(b).  For $\delta>\delta_c$ we deduce a LT as function of $w$, see Fig.~\ref{fig:1}(d). We now apply numerical renormalization group (NRG) calculations to demonstrate 
these signatures more quantitatively.

\begin{figure}
\includegraphics[width=0.95\linewidth]{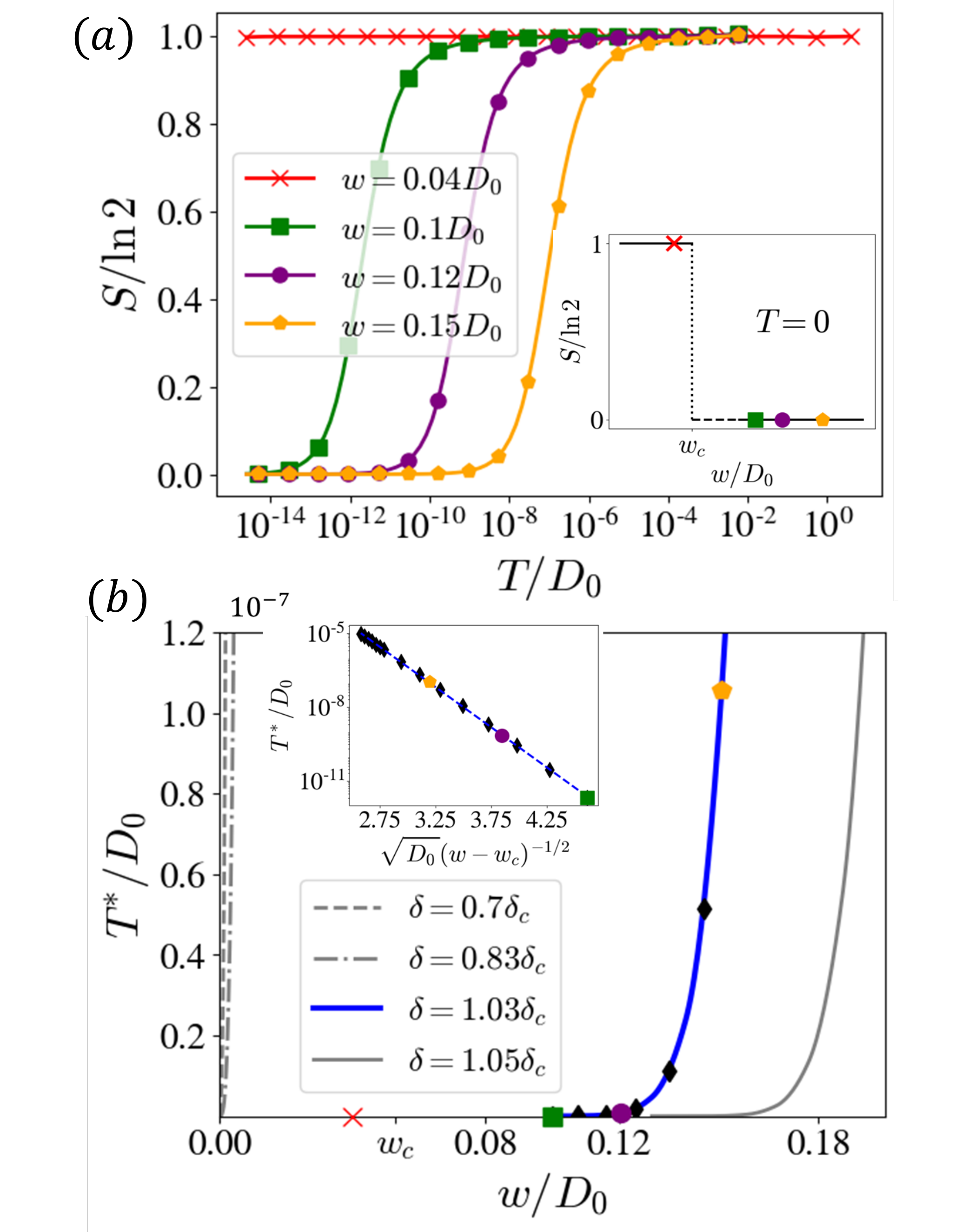}
\caption{(a) NRG results for  the entropy versus $T$ for various $w$ and for $\delta=1.03 \delta_c$. For $w>w_c$ the entropy drops to zero at low temperatures, but for $w<w_c$ it remains $\ln{2}$ down to $T=0$ as exemplified by the sudden change in the  entropy at $T=0$ at $w=w_c$, depicted in the inset.
(b) Crossover temperature $T^*$ for various $\delta$ as a function of $w$. The blue thick line corresponds to the parameter $\delta=1.03 \delta_c$ of (a), with the 4 colored markers denoting the crossover temperatures for the 4 different curves in (a). Inset: $T^*/D_0$ as a function of $D_0/(w-w_c)$, demonstrating the dependence $T^*\propto \exp(-const\times(w-w_c)^{-1/2})$, from which we extract $w_c$. }

\label{fig:6}. 
\end{figure}


\emph{NRG results---} The spinless DD is described by
\be
\label{eq:syslattuceDD}
	H_{DD}= -\mu(a^\dagger a + b^\dagger b)+ \Delta (a^{\dagger} a - b^{\dagger} b) - w (a^\dagger b+h.c.)+U a^\dagger a b^\dagger b, 
	\ee
 where $\mu$ and $\Delta$ denote, respectively, the DD 
 chemical potential and asymmetry. 
 As finite asymmetry smears the LT~\cite{SM}, we focus here on the symmetric case $\Delta=0$.  Here $a(b)$ annihilates an electron in QD $A(B)$, $\hat{N}_A=a^\dagger a$ and we define the DD occupancy $N \equiv \langle a^\dagger a + b^\dagger b \rangle$. $\mu$ is used to continuously switch from the empty regime $N=0$ to the singly occupied regime $N=1$. We assume $U \to + \infty$ to exclude double occupancy. We ignore real spin, assuming  that a particular electron spin is being trapped in the DD.

Our NRG calculations solve a fermionic lattice model corresponding to Eq.~(\ref{eq:QPC}), which is also equivalent at low energy to the effective Hamiltonian Eq.~(\ref{eq:O}) and hence reproduces its critical properties. It consists~\cite{SM} of a fermionic semi-infinite tight binding chain interacting near the origin with the DD. The  interaction term is selected~\cite{SM} to yield  the desired $N_A$-dependent phase shift $\delta$. We compute the entropy, the total charge of the DD, and the many-body energy levels.


Fig.~\ref{fig:6}(a) shows the entropy $S(T)$ in the singly occupied regime. We consider the interesting case with $\delta> \delta_c$. For $w>w_c$, $S(T)$ displays a drop by $\ln{2}$ below a characteristic energy scale $T^{*}$ (defined as $S(T^{*}) = \frac{1}{2}\ln{2}$). As displayed in Fig.~\ref{fig:6}(b) by the thick blue curve, upon decreasing $w$, $T^*$ decreases and eventually vanishes at $w=w_c(\delta)$. The precise form of the vanishing of $T^*$ is shown in the inset, demonstrating the scaling behavior expected near the KT transition. The resulting phase diagram in Fig.~\ref{fig:6}(b), which is plotted for few values of $\delta$, has the structure of Fig.~\ref{fig:1}(c) for $\delta < \delta_c$ and Fig.~\ref{fig:1}(d) for $\delta >\delta_c$. 
In particular, for $\delta < \delta_c$, $T^{*}$ only vanished at $w=0$.
Thus the LT features a discontinuous change of entropy at $T \to 0$ as a function of  $w$, see inset of Fig.~\ref{fig:6}(a). 

\emph{Entropy from Maxwell relations---} 
  Experimentally, changes in the entropy upon varying the DD chemical potential $\mu: \mu_1 \to \mu_2$ are accessible via the Maxwell relation~\cite{hartman2018direct,child2022entropy,child2022robust,sela2019detecting,han2022fractional}
 \be
 \Delta S_{\mu_1\to\mu_2} = \int_{\mu_1}^{\mu_2} \frac{dN(\mu)}{dT}d\mu.
 \label{eq:maxwell}
\ee
Namely, by using the QPC as a charge detector, one measures the differential charging curve $dN/dT$ upon varying $\mu$ from the empty to the singly occupied regime. Since the entropy vanishes in the empty-DD regime, we obtain the entropy of the spin-boson model described by the singly occupied regime from this integral.

\begin{figure}
\includegraphics[width=0.95\linewidth]{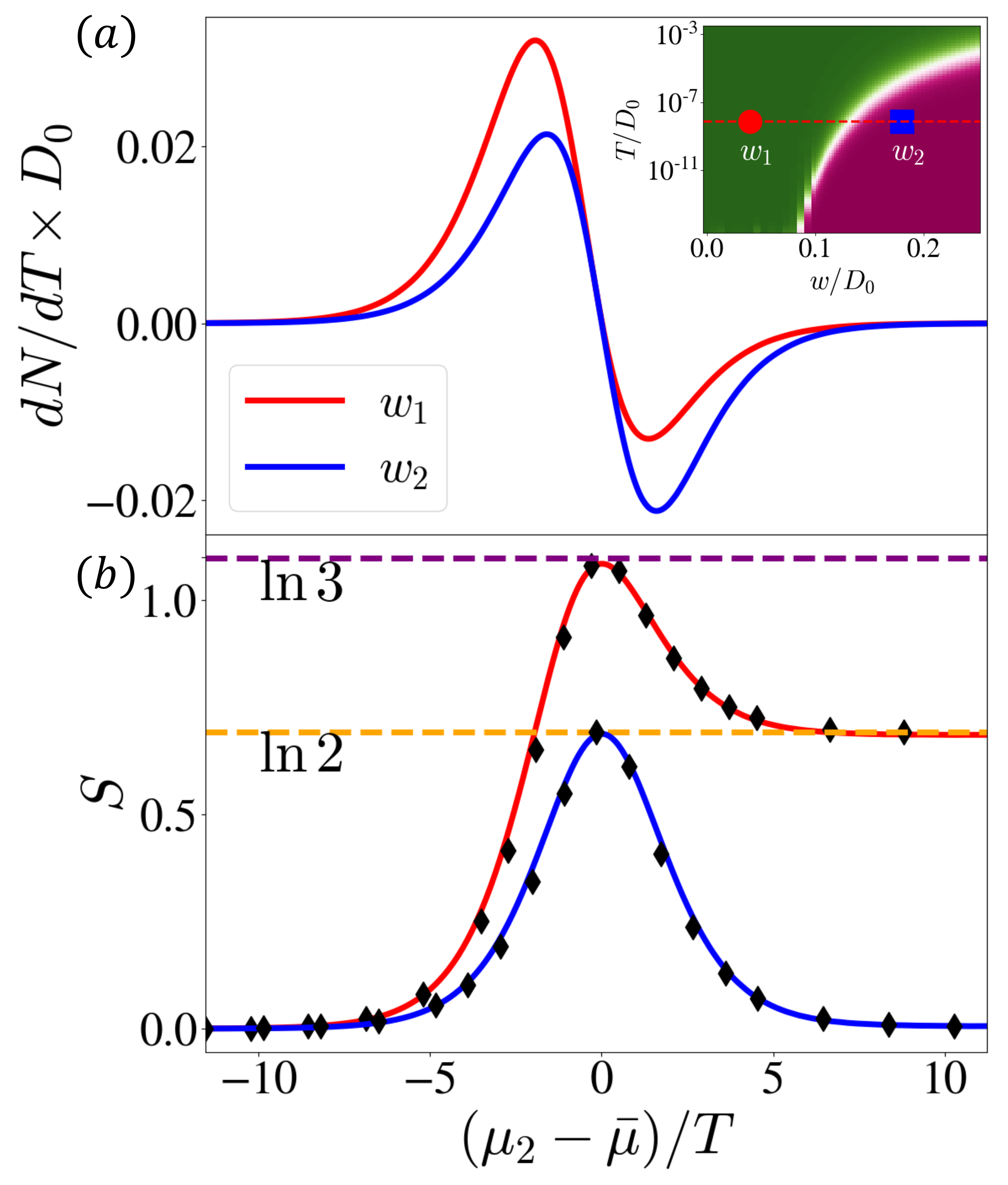}
\caption{
(a) $dN/dT$ for $\{w_1/D_0,w_2/D_0\}=\{0.03, 0.17\}$ as function of the DD chemical potential for $T/D_0=7.5\times 10^{-9}$. Red (blue) lines describe the decoherence (coherence) phase, corresponding to the green (magenta) region as depicted in the inset.
(b) Entropy  $S(\mu_2)$ obtained by integrating  $d N/d T$, see Eq.~(\ref{eq:maxwell}). We shifted the $\mu$ axis such that  peaks  occur at the origin. The blue curves for $w>w_c$ agree with the entropy of a decoupled DD, while for $w<w_c$ the system is driven to the incoherent phase. As a comparison, the black diamonds show the entropy obtained directly from NRG.}\label{fig:3}
\end{figure}

Fig.~\ref{fig:3} displays $dN/dT$ as calculated from NRG (top panel) and the entropy change $S(\mu_2)$ as obtained by integration from $\mu_1 = -\infty$ to $\mu_2$  (lower panel), 
 for two different values of $w$, as shown in the inset of Fig.~\ref{fig:3}, at a fixed temperature.  In the absence of the QPC, the ground state of the DD  is unique (e.g. the symmetric state), and thus the entropy increases as a function of $\mu$, from zero, in the empty-DD regime, to $\ln{2}$, when the empty and singly occupied states are degenerate and then decrease back to zero in the singly occupied regime. This is observed when $w>w_c$ (blue curves). However, once $w$ becomes smaller than $w_c$ (or $\delta$ becomes larger than $\delta_c$ for this value of $w$), the behavior changes abruptly. Now, due to the loss of coherence between the two QDs, the two singly occupied states $\ket{N=1, N_A=0,1}$ are degenerate, resulting in the increase of the entropy to $\ln{3}$ before dropping to $\ln{2}$ for the singly occupied state. 

\emph{Conductance jump---} For a 2D superfluid, the Kosterlitz-Thouless RG equations result in a universal jump in the superfluid density~\cite{nelson1977universal}. What is then the corresponding discontinuous quantity in our system? 

The DD creates a different single-particle scattering potential on the QPC for each value of $N_A$. We can use the Landauer formula, which gives the conductance at $T=0$
\be
\label{eq:Landauer}
G_{N_A}=\frac{2e^2}{h}\sum_n \cos^2 (\delta_{N_A}^{(e,n)}-\delta_{N_A}^{(o,n)}).
\ee
So a discontinuity in $\delta$  yields a discontinuity in $G_{N_A}$.
To be concrete, consider the 
model
\be
\label{eq:V}
H_{QPC}=-\frac{\hbar^2}{2m}(\partial_x^2+\partial_y^2) +\frac{m \omega^2 y^2}{2}-\frac{\hbar \omega}{2}+\frac{V_{0}  |0\rangle \langle 0 |+V_{1}  |1\rangle \langle 1 |}{\cosh^2 (x/a)}.
\ee
The Fermi momentum $k_{F,n}$ of the $n-$th transverse mode satisfies $E_F=\hbar^2 k_{F,n}^2/2m+n \hbar \omega$ with $n=0,1,\dots,\lfloor E_F/\omega\rfloor$. We let $V_0=V_g$ and $V_1=V_g+\Delta V$, with fixed $\Delta V$ characterizing the electrostatic interaction and a parameter $V_g$ tunable using a gate voltage.

\begin{figure}
    \centering
    \includegraphics[width=0.95\linewidth]{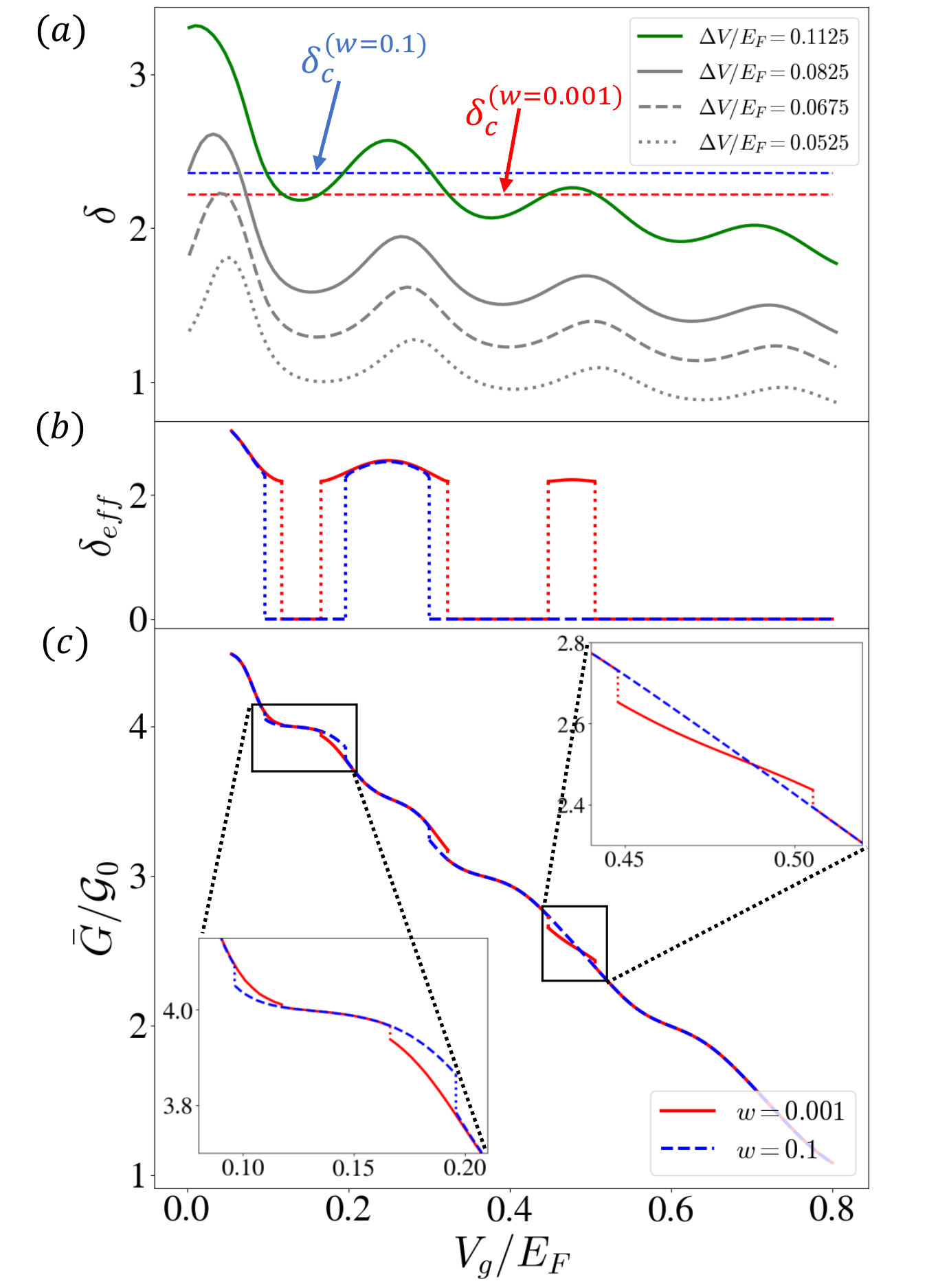}
    \caption{(a) Phase shift $\delta$ for various electrostatic couplings
    $\Delta V$, versus a gate voltage controlling the potential barrier ($E_F/\omega=4.3$ and $\hbar^2/(2ma^2)/E_F=100$). Dashed horizontal lines denote $\delta_c^{(w)}$ for two values of $w$.   (b) Renormalized phase shift $\delta_{eff}$ for $\Delta V/E_F=0.1125$ and $w/D_0=0.001$ (red) or $w/D_0=0.1$ (blue). Whenever $\delta$ crosses $\delta_c^{(w)}$ there is a discontinuous jump in $\delta_{eff}$.  (c) Corresponding conductance jumps. }
    \label{fig:my_label}
\end{figure}

For each mode, one can analytically compute~\cite{cevik2016resonances} the even and odd phase shifts, and thus obtain $\delta$. In Fig.~\ref{fig:my_label}(a) we  plot the calculated $\delta$ versus $V_g$ for a constant $\Delta V$ for selected parameters corresponding to 5 transverse modes. Then $\delta$ displays peaks approximately when a transverse mode becomes reflecting. $\delta_c^{(w)}$ is marked by dashed lines for two values of $w$. In each case, we can see regions where $\delta > \delta_c^{(w)}$ are achieved for large enough $\Delta V$.

From Fig.~\ref{fig:1}(b), one can see that upon increasing $\delta$,  as obtained by continuously varying $V_g$, when the condition $\delta > \delta_c^{(w)}$ is met, the effective interaction $\delta(\ell \to \infty) \equiv \delta_{\rm{eff}}$ suddenly jumps from 0 to $\delta_c=\pi/\sqrt{2}$. In Fig.~\ref{fig:my_label}(b) we plot $\delta_{\rm{eff}}$ as  extracted from the NRG finite-size spectrum~\cite{SM}, indeed demonstrating these sharp jumps.

For either  the coherence or decoherence fixed points with $\delta \to 0$ or $w \to 0$, respectively, one can recombine the two terms $H=H_{{\rm{eff}}}(\phi) + \sum_{i=2}^{i_{max}} H[\phi']$ 
by replacing $\delta \to \delta_{\rm{eff}}$. 
Returning to the original basis $\{\phi_i\}$, one can read off the even and odd phase shifts in each 
channel, 
\be
\label{eq:deltaieff}
\left( \delta^{(i)}_{0,1}\right)_{{\rm{eff}}}=
\bar{\delta}_i
\pm 
\delta_i
\frac{\delta_{{\rm{eff}}}}{\delta}. 
\ee
Substituting Eq.~(\ref{eq:deltaieff}) in the expression (\ref{eq:Landauer}) for the  conductance, we see that when $\delta_{\rm{eff}}$ changes discontinuously across the LT, so do both $G_{0}$ and $G_1$. In Fig.~\ref{fig:my_label}(c) we plot the average conductance $\bar{G}=(G_0+G_1)/2$ at $T=0$. We see that it displays discontinuities  precisely when the LT is crossed for each value of $w$ (in Ref.~\onlinecite{SM} we show that a similar discontinuity can occur for a fixed $V_g$ as a function of $w$). Thus, the LT of the KT type can be inferred from the conductance itself. 

\emph{Summary---} Recent experiments demonstrated the ability to measure entropy changes in mesoscopic systems by coupling them to charge detectors. Here, we demonstrate that even at thermal equilibrium the charge detector may strongly affect the system and drive an environment-induced localization transition. The resulting entropy change describes the process of a quantum measurement of a state as it is being measured by an environment. Relating this entropy change due to measurement of a subsystem to entanglement entropy between the two subsystems is left for future work~\cite{macieszczak2019coherence,ma2021symmetric,PhysRevLett.130.136201}.

\begin{acknowledgements}
    \emph{Acknowledgements---} We gratefully acknowledge support from the European Research Council (ERC) under the European Unions Horizon 2020 research and innovation programme under grant agreement
No. 951541. ARO (W911NF-20-1-0013) and the Israel Science Foundation grant numbers 154/19. We thank discussions with Josh Folk, Andrew Mitchell, and Sarath Sankar. 
\end{acknowledgements}

\bibliography{QIbasedNonAbelian}

\end{document}


\title{Identifying an environment-induced localization transition from entropy and conductance: Supplemental Material
}
	\author{Zhanyu Ma}
	\affiliation{School of Physics and Astronomy, Tel Aviv University, Tel Aviv 6997801, Israel}

	\author{Cheolhee Han}
	\affiliation{School of Physics and Astronomy, Tel Aviv University, Tel Aviv 6997801, Israel}
	
	\author{Yigal Meir}
\affiliation{Department of Physics, Ben-Gurion University of the Negev, Beer-Sheva, 84105 Israel}
		\author{Eran Sela}
	\affiliation{School of Physics and Astronomy, Tel Aviv University, Tel Aviv 6997801, Israel}

\maketitle

This supplementary material includes (i) details on our numerical renormalization group (NRG) calculations, (ii) NRG results on the double-dot (DD)  and the resonant level model (RLM), (iii) an analysis of the asymmetry in the DD model together with asymmetry in the QPC, and (iv) few details on our effective field theory. 
\section{Details on our NRG calculations}
\label{se:nrgdetails}
\begin{figure}
\centering
\includegraphics[width=.8\columnwidth]{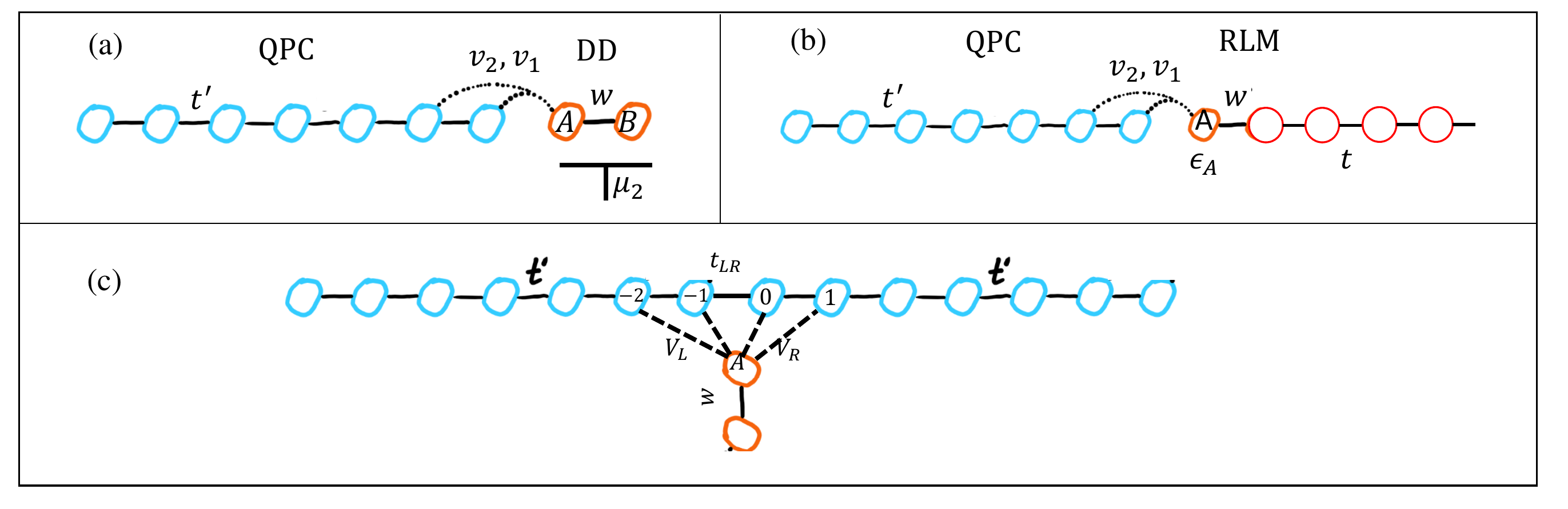}
\caption{Schematics of the tight-binding chain used in our NRG simulations for our (a) DD model, (b) RLM, and (c) off-diagonal scattering model. Dashed lines denote interactions and solid lines denote hopping amplitudes.}
\label{fig:tightbindingchain}
\end{figure}

In this section, we detail our NRG analysis.
We start with a semi-infinite tight-binding model with a local potential that depends on $N_A$, the number of electrons on QD $A$. The Hamiltonian is
$H = H^{(f)}_{QPC} + H_{sys}$, where the system's Hamiltonian $H_{sys}$ is specified below, and the QPC is described by 
\be
\label{eq:NRGtightbinding}
H^{(f)}_{QPC}=-t'\sum_{i=1}^{\infty} (f^\dagger_i f_{i+1}+h.c.) -  \sum_{i=1}^{i_{max}}v_i (f_i^{\dag}f_i - \frac{1}{2})(\hat{N}_A - \frac{1}{2}),
\ee
see Fig.~\ref{fig:tightbindingchain}. 

For each value of $N_A=0,1$, $H^{(f)}_{QPC}$ describes a non-interacting scattering problem.
Incoming waves pick up from the potential a scattering phase shift which depends on $N_A$. While the QPC in Fig.~1 in the main text contains source and drain leads, allowing to define its conductance, the only important parameter is the total phase shift difference over all parity quantum numbers, transverse modes and spin. Thus Eq.~(\ref{eq:NRGtightbinding}) is an economic model taking into account only a single mode (and no even/odd parity quantum numbers).

By fixing $N_A$, one can solve  analytically for the phase shift, as follows. In our tight-binding  model, the phase shift switches sign as $N_A=1 \to 0$, hence $\delta =| \delta_{N_A=1}|$. Solving the Schr\"{o}dinger equation, using the asymptotic wavefunction $\psi_j\propto \sin(kj + \delta(k))$, we find
\be
\delta(k) = \tan^{-1}\frac{2\alpha\cos{k}\sin{2k} - \frac{1}{2}\sin{2k} + \alpha\sin{k}}{2\alpha^2 - (2\alpha\cos{k}\cos{2k} - \frac{1}{2}\cos{2k}+\alpha\cos{k})},
\ee
where we have set $v_1=v_2=V$, and $\alpha = t'/V$. $\delta$ varies on an energy scale set by the bandwidth $t'$, thus $\delta$ can be approximated by a constant for  $T \ll  t'$. 
For the half filled chain, we use $\delta=\delta(k_F=\pi/2) = \tan^{-1}\frac{2\alpha}{4\alpha^2 -  1}$. The phase shift $\delta$ is the dimensionless coupling constant of the problem. Given this analytic relation between $\delta$ and $\{ v_i\}$, we plot various quantities (like entropy) as function of $\delta$, and not $\{ v_i \}$.

Unless specified otherwise we used an interaction range $i_{max}=2$,  $v_1=v_2=V$, the bandwidth of QPC is fixed to unity, $t'=1$ (thus all other energies are measured in this unit). The NRG discretization parameter is fixed to $\Lambda=2$ and we keep 800 states per iteration. 
The entropy calculated from NRG is often referred to as the impurity entropy $S_{imp}$, being the difference between the entropy of the full system $H_{QPC}^{(f)}+H_{sys}$ and the one without the ``impurity", namely $H_{QPC}^{(f)}$.

\section{$\delta_{{\rm{eff}}}$ from NRG}
\label{se:NRGdelta}
\begin{table}[h]
\begin{tabular}{@{}lllll@{}}
\toprule
\multicolumn{1}{c|}{\multirow{2}{*}{}} & \multicolumn{2}{c|}{$w=0.0001$} & \multicolumn{2}{c}{$w=0.2$} \\ \cmidrule(l){2-5} 
\multicolumn{1}{c|}{}                  & $\tilde{E}_{NRG}/C$            & \multicolumn{1}{l|}{$LE_{fp}/(2\pi)$}         & $\tilde{E}_{NRG}/C$          & $LE_{fp}/(2\pi)$       \\ \cmidrule(r){1-1}
\multicolumn{1}{l|}{state 1}                               & \multicolumn{1}{l|}{0}              & \multicolumn{1}{l|}{0}           & \multicolumn{1}{l|}{0}            & {0}          \\ \hline
\multicolumn{1}{l|}{state 2}                               & \multicolumn{1}{l|}{0}              & \multicolumn{1}{l|}{0}           & \multicolumn{1}{l|}{0.5}          & {0.5}        \\ \hline
\multicolumn{1}{l|}{state 3}                               & \multicolumn{1}{l|}{0.2364}         & \multicolumn{1}{l|}{$\frac{\delta_{{\rm{eff}}}}{\pi}-\frac{1}{2}=0.2364$}        & \multicolumn{1}{l|}{0.5}          & {0.5}        \\ \hline
\multicolumn{1}{l|}{state 4}                               & \multicolumn{1}{l|}{0.2364}         & \multicolumn{1}{l|}{$\frac{\delta_{{\rm{eff}}}}{\pi}-\frac{1}{2}=0.2364$}          & \multicolumn{1}{l|}{1}            & {1}          \\ \hline
\multicolumn{1}{l|}{state 5}                               & \multicolumn{1}{l|}{0.7636}         & \multicolumn{1}{l|}{$\frac{3}{2} - \frac{\delta_{{\rm{eff}}}}{\pi}=0.7636$}         & \multicolumn{1}{l|}{1.507}        & {1.5}        \\ \hline
\multicolumn{1}{l|}{state 6}                               & \multicolumn{1}{l|}{0.7636}         & \multicolumn{1}{l|}{$\frac{3}{2} - \frac{\delta_{{\rm{eff}}}}{\pi}=0.7636$}         & \multicolumn{1}{l|}{1.507}        & {1.5}        \\ \hline
\multicolumn{1}{l|}{state 7}                               & \multicolumn{1}{l|}{1}              & \multicolumn{1}{l|}{1}           & \multicolumn{1}{l|}{2.007}        & {2}          \\ \hline
\multicolumn{1}{l|}{state 8}                               & \multicolumn{1}{l|}{1}              & \multicolumn{1}{l|}{1}           & \multicolumn{1}{l|}{2.007}        & {2}          \\ \hline
\multicolumn{1}{l|}{state 9}                               & \multicolumn{1}{l|}{1.2278}         & \multicolumn{1}{l|}{$\frac{\delta_{{\rm{eff}}}}{\pi} + 1=1.2364$}         & \multicolumn{1}{l|}{2.007}        & {2}          \\ \hline
\multicolumn{1}{l|}{state 10}                              & \multicolumn{1}{l|}{1.2278}         & \multicolumn{1}{l|}{$\frac{\delta_{{\rm{eff}}}}{\pi} + 1=1.2364$}        & \multicolumn{1}{l|}{2.007}        & {2}          \\ \bottomrule
\end{tabular}
\caption{\label{deleff} Comparison of the NRG finite size spectrum and the fixed point spectrum Eq.~(\ref{eq:fps}). We fix $\delta=1.03\delta_c$ and find $C=1.31$. For $w=0.0001$, the system flows to the decoherence fixed point, with doubly degenerate spectrum. We find $\delta_{{\rm{eff}}} = 0.7364$. For $w=0.2$, the system flows to the coherence fixed point, with $\delta_{{\rm{eff}}}=0$.}
\end{table}
We  extract $\delta_{{\rm{eff}}}$ (see Fig.~4 in the main text) from the NRG finite size spectrum. The underlying principle is that the NRG fixed point is a boundary conformal field theory, whose energy spectrum is given by
\be
H_{fp} = \frac{2\pi}{L} \sum_{q \in \mathbb{Z} ~{\rm{or}} ~\mathbb{Z}+\frac{1}{2}} (q - \frac{\delta_{{\rm{eff}}}}{\pi}) c_q^{\dag}c_q,
\label{eq:fps}
\ee
for a system of size $L$. Here $q$ takes integer (half-integer) values for even (odd) $L$, see Refs.~\cite{borda20034, hofstetter2004singlet}. So by fitting the NRG fixed point spectrum with Eq.~(\ref{eq:fps}), 
\be
\lim_{N\to\infty} \tilde{H}_N = C \frac{L}{2\pi} H_{fp},
\ee
we can extract $\delta_{{\rm{eff}}}$. Here $\tilde{H}_N$ is the rescaled Hamiltonian in NRG, $C$ is a $\Lambda$-dependent constant to be determined numerically, with $\Lambda = 2$ in our NRG calculation. In Table~\ref{deleff}, we list 10 lowest eigenstates of the NRG fixed point spectrum, and the corresponding fitting based on the boundary conformal field theory Eq.~(\ref{eq:fps}), for $w=0.0001<w_c$ and $w=0.2>w_c$, separately. We see for $w=0.0001<w_c$ that the spectrum is doubly degenerate, reflecting the localization of the electron in the DD.

\section{NRG results for the DD and RLM models}
\label{se:systems}
In the main text, we studied a DD coupled to a QPC. However, the LT happens more generally. In this section, we  exemplify this for the RLM coupled to a QPC. 

Generally, the system's Hamiltonian $H_{sys}$ contains two subsystems $A$ and $B$ connected by a tunneling term $H_{tun}=H_{tun}^{(+)} + H_{tun}^{(-)}$ where $H_{tun}^{(+)}(H_{tun}^{(-)})$ increases (decreases) $N_A$ by one. We  consider only two possible charge states for subsystem $A$, $N_A=0,1$, represented by a Pauli matrix $\sigma^z = \pm 1$.

Let us assume that for the isolated system, the tunneling $w$ has a bare scaling dimension  $x_{tun}$. This means that the tunneling amplitude satisfies the RG equation $dw/d\ell = w(1 - x_{tun})$.
Coupling the system to the QPC via $H_{{\rm{eff}}}(\phi)$ in the main text, the RG equation becomes
\be
\label{eq:RG1}
\frac{dw}{d\ell} = w(1 - x_{tun} - x_b),~~~x_b=2(\delta/\pi)^2.
\ee
We denote the RG energy scale by $D$ and the bare scale by $D_0=t'$ as set by the bandwidth of the QPC, such that $\ell =\log(D_0/D)$. Then the RG equation is solved by $w(D)=w(D_0) \left(\frac{D_0}{D} \right)^{1-x_{tun}-x}$. Assuming that the bare coupling $w(D_0)\equiv w$ is small, setting $w(T^*)=1$, leads to the energy scale
\be
\label{eq:kttransisiotn}
T^* =D_0  w^{\frac{1}{1-x_{tun}-x}} \approx D_0  w^{\frac{\pi^2}{4 \delta_c (\delta_c - \delta)}},
\ee
which  vanishes at the LT. In the second equality, we expanded $1-x_{tun}-x$ to first order in $\delta$ near $\delta_c$.

\subsection{Double dot model}

The DD model has a Hamiltonian $H_{sys}=H_{DD}= -\mu(a^\dagger a + b^\dagger b)+ \Delta (a^{\dagger} a - b^{\dagger} b) - w (a^\dagger b+h.c.)+U a^\dagger a b^\dagger b$, reproduced from the main text for clarity. For zero asymmetry $\Delta=0$, the two charge states are degenerate. The tunneling operator $\propto w$ that connects them has scaling dimension $x_{tun}=0$, see Table~\ref{Tab:table1}. From the RG equation Eq.~(\ref{eq:RG1}) the LT $dw/d \ell=0$ occurs at $\delta_c=\pi/\sqrt{2}$, see Table~\ref{Tab:table1}.

\begin{figure}
\includegraphics[width=.7\columnwidth]{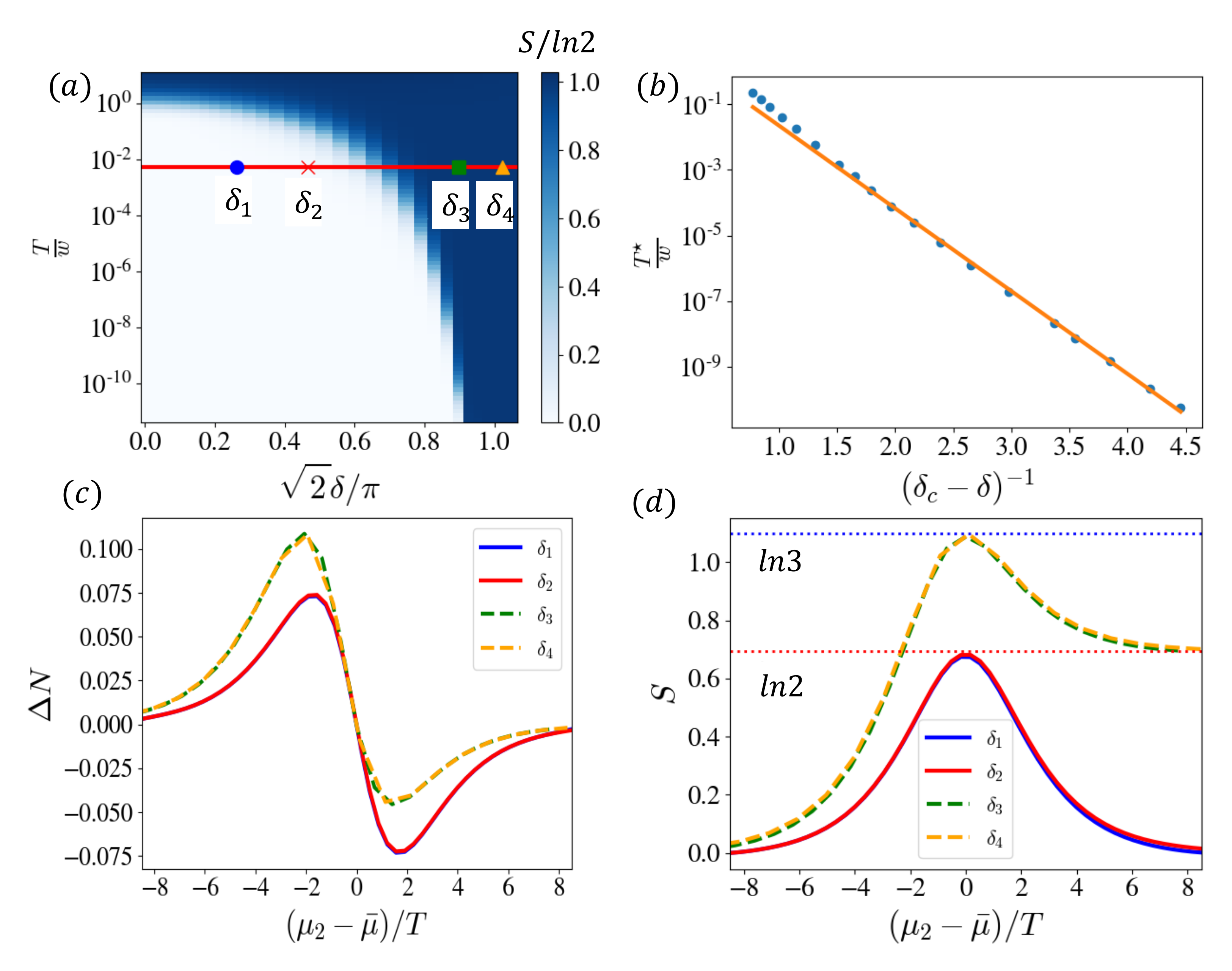}
    \caption{
DD results.  (a) Impurity entropy  as a function of normalized temperature $T/w$ and normalized phase shift $\delta/\delta_c$. We use $w/D_0 =0.001$.
(b) The crossover scale $T^*$ follows the scaling form of Eq.~(\ref{eq:kttransisiotn}) with $\delta_c = \pi/\sqrt{2}$. (c,d) Same as Fig.~3 in the main text, but for the specified values of $\delta_1, \dots , \delta_4$ in (a). }  
    \label{fig:my_label1}
\end{figure}

We display in Fig.~\ref{fig:my_label1}(a) the entropy versus temperature and $\delta$; the scaling of $T^*$ is shown in Fig.~\ref{fig:my_label1}(b). Figs.~\ref{fig:my_label1}(c-d) are similar to Figs.~3 in the main text, but for the specified values of $\delta$.

\subsection{Resonant level model}
Now we consider a RLM $H_{sys}=H_{RLM}$, given by the Hamiltonian
\bea
\label{eq:syslattuce}
H_{RLM}= \epsilon_a a^\dagger a -t \sum_{i=1}^\infty (c^\dagger_{i} c_{i+1}+h.c.) +w (a^\dagger c_1 +h.c),
\eea
with $\hat{N}_A=a^\dagger a$, i.e. instead of a second dot in the DD system, dot $A$ is now coupled to a second quantum wire, described by a non-interacting tight-binding model, see Fig.~\ref{fig:tightbindingchain}(b). 
When $\epsilon_a=0$ the two states with $N_A=0,1$ are degenerate. 
The main difference compared to the DD case is that the tunneling operator $w$ includes a lead operator $c_1$ which has scaling dimension $x_{tun}=1/2$, see Table~\ref{Tab:table1}. Then the transition occurs when $\delta=\pi/2<\pi/\sqrt{2}$, namely a weaker interaction with the QPC is needed to drive the LT, see Table~\ref{Tab:table1}. In our NRG calculation, we assume the bandwidth of the lead is also fixed to unity, $t=D_0=1$.

\begin{figure}
    \centering
    \includegraphics[width=.7\columnwidth]{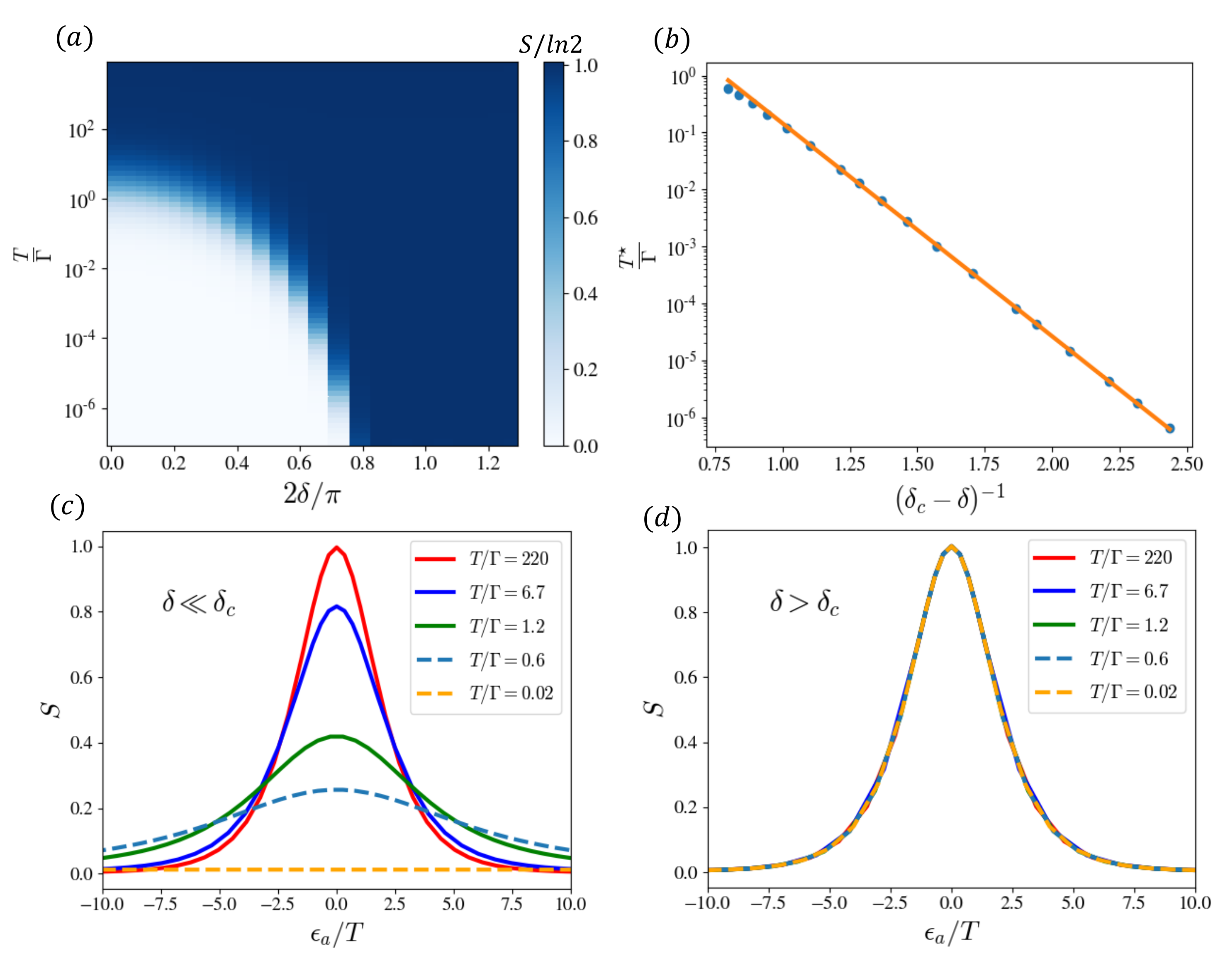}
    \caption{NRG results for the resonant level model.
In panel (a), we display the impurity entropy in units of $\ln{2}$ as a function of normalized temperature $T/\Gamma$ and normalized phase shift $\delta/\delta_c$. Notice the similarity between panel (a) and the phase diagram for the DD.
In panel (b), we show that $T^*$ follows Eq.~(\ref{eq:kttransisiotn}) with $\delta_c = \pi/2$.
In panels (c) and (d) we show the dependence of $S_{imp}$ on the level position $\epsilon_a$. 
In panel (c), there is no interaction with the QPC, $\delta = 0$ and in panel (d) we are in the strong interaction regime, $\delta = \frac{\pi}{2}\times 1.2 > \delta_c = \frac{\pi}{2}$. The fact that in panel (d) all curves  collapse is a manifestation of the LT.}
    \label{fig:my_label}
\end{figure}

Our NRG results for the RLM model are summarized in Fig.~\ref{fig:my_label}. Fig.~\ref{fig:my_label}(a) shows the  entropy versus temperature and interaction strength $\delta$. At $\delta = 0$, the entropy $S(T)$ displays a drop by $\ln{2}$ as $T$ is reduced below $\Gamma=w^2 /t$ in accordance with the behavior of a decoupled RLM. The $\ln{2}$ entropy  at $T \gg \Gamma$ represents the two possible occupations of the  level. At low temperature, $T \ll \Gamma$ the  level is hybridized with the lead and this entropy is quenched. As $\delta$ increases, we observe a decreasing energy scale which eventually vanishes at $\delta=\delta_c$ according to $T^* \sim w^{\frac{const}{\delta_c-\delta}}$, see Fig.~\ref{fig:my_label}(b). 

The $\ln{2}$ entropy for $T \gg \Gamma$ is quenched upon shifting the level away from resonance, as seen in Fig.~\ref{fig:my_label}(d). 
The peak is suppressed at $T \ll \Gamma$.  However, 
as seen in Figs.~\ref{fig:my_label}(d), for $\delta > \delta_c$, in the localized phase, all curves corresponding to vastly different temperatures collapse to the same curve because the tunneling term $\Gamma$ is effectively absent.

\begin{table}[h!]
\caption{$x_{tun}$ and $\delta_c$ for the double dot and resonant level models.}
  \begin{center}
    \label{tab:table1}
    \begin{tabular}{l|c|r} 
      & $\boldsymbol{x_{tun}}$ & $\boldsymbol{\delta_c}$\\
      \hline
      DD & $0$ & $\pi / \sqrt{2}$\\
      RLM & $1/2$ & $\pi / 2$
    \end{tabular}
  \end{center}
  \label{Tab:table1}
\end{table}

\subsection{Conductance jump tuned by $w$}
In the main text, we illustrated the conductance jump observed upon crossing the LT as a function of $\delta$. Here, we demonstrate that the LT in the conductance can also be seen as a function of $w$.

\begin{figure}
    \centering
    \includegraphics[width=.7\linewidth]{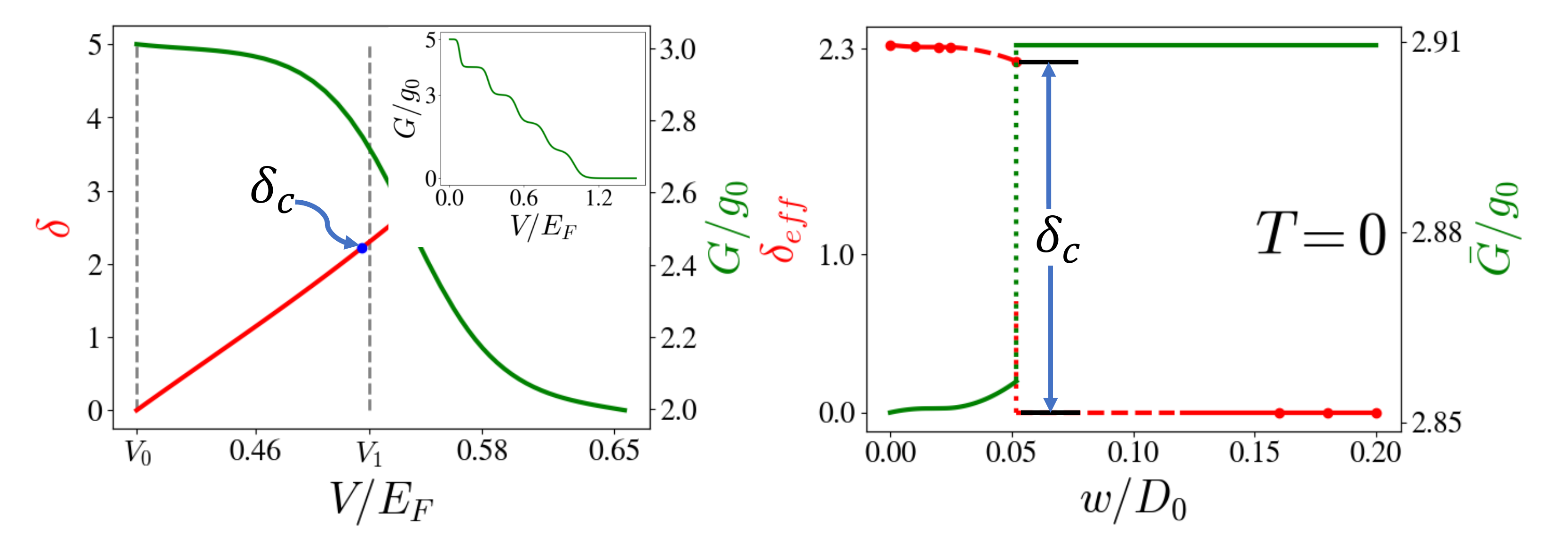}
    \caption{(a) Inset: conductance $G$ versus $V$ for the QPC model Eq.~(10) of the main text for  $\frac{E_F}{\hbar \omega}=4.3$ corresponding to $5$ transverse modes, and $\frac{\hbar^2/(2ma^2)}{E_F}=100$.
    Main panel: one conductance step, and  $\delta$ for fixed $V_0$ as function of $V_1$. We mark $\delta_c$ by a blue dot and select $V_1$ as in the corresponding vertical dashed line. (b) Effective interaction $\delta_{{\rm{eff}}}$ and average conductance $\bar{G}$ as function of $w$. $\delta_{{\rm{eff}}}$ displays a universal jump of $\pi/\sqrt{2}$ at $w_c$. }
    \label{fig:Gstep}
\end{figure}

In the inset of Fig.~\ref{fig:Gstep}(a) we plot the conductance Eq.~(9) in the main text versus $V$, the coefficient of the $1/\cosh^2(x/a)$ potential, for  $\frac{E_F}{\hbar \omega}=4.3$ corresponding to $5$ spinful transverse modes, and $\frac{\hbar^2/(2ma^2)}{E_F}=100$.

Now we let $V$ depend on $N_A=0,1$, $V=V_{N_A}$,  select a value $V_0/E_F=0.4$, and plot in Fig.~\ref{fig:Gstep}(a) both the calculated $\delta$ and $G$  as function of $V_1$. We then select $V_1$ such that $\delta > \delta_c$ (vertical dashed line).

As explained above, we compute $\delta_{{\rm{eff}}}$ from NRG. Fig.~\ref{fig:Gstep}(b) depicts   $\delta_{\rm{eff}}$, which is zero for $w>w_c(T)$, while for $w<w_c(T)$  $\delta_{\rm{eff}}$ is finite, and coincides with the bare $\delta$ (of Fig.~\ref{fig:Gstep}(a)) for  $w \to 0$. The jump of $\delta_{\rm{eff}}$ at $w=w_c$ yields $\delta_c$.

Importantly, the QPC acts as a charge sensor due to the dependence of the phase shifts on $N_A=0,1$. 
However, for $w>w_c$, $\delta \to 0$, we have $G_0=G_1$ and the visibility of the charge detector vanishes. For $w<w_c$ we see from the flow diagram Fig.~1(a) of the main text that $w \to 0$, meaning
that the conductance is either given by $G_0$ or $G_1$,  for the two values of $N_A$ which no longer fluctuates. 

While $G_0$ and $G_1$ can be measured independently from the telegraphic noise as an electron  hops into and out of QD $A$~\cite{field1993measurements}, we consider the average conductance $\bar{G}=(G_0+G_1)/2$. 
As seen in Fig.~\ref{fig:Gstep}(b), the universal jump in $\delta_{\rm{eff}}$ results in a non-universal jump in $\bar{G}$ which depends on all the phase shifts. But, its location at $w=w_c$ is indicative of the LT. 
\section{Effects of asymmetry $\Delta$ on the DD}
\label{se:asy}
Here we explore the effects of asymmetry $\Delta$ in the DD model. For finite $\Delta$ the DD  coupled to a QPC maps to the spin-boson model with an additional longitudinal magnetic field $\Delta \sigma^z$ breaking the degeneracy between the two eigenstates of $\sigma^z$. Hence the LT does not exist as a quantum phase transition.

\begin{figure}
    \centering
    \includegraphics[width=.8\linewidth]{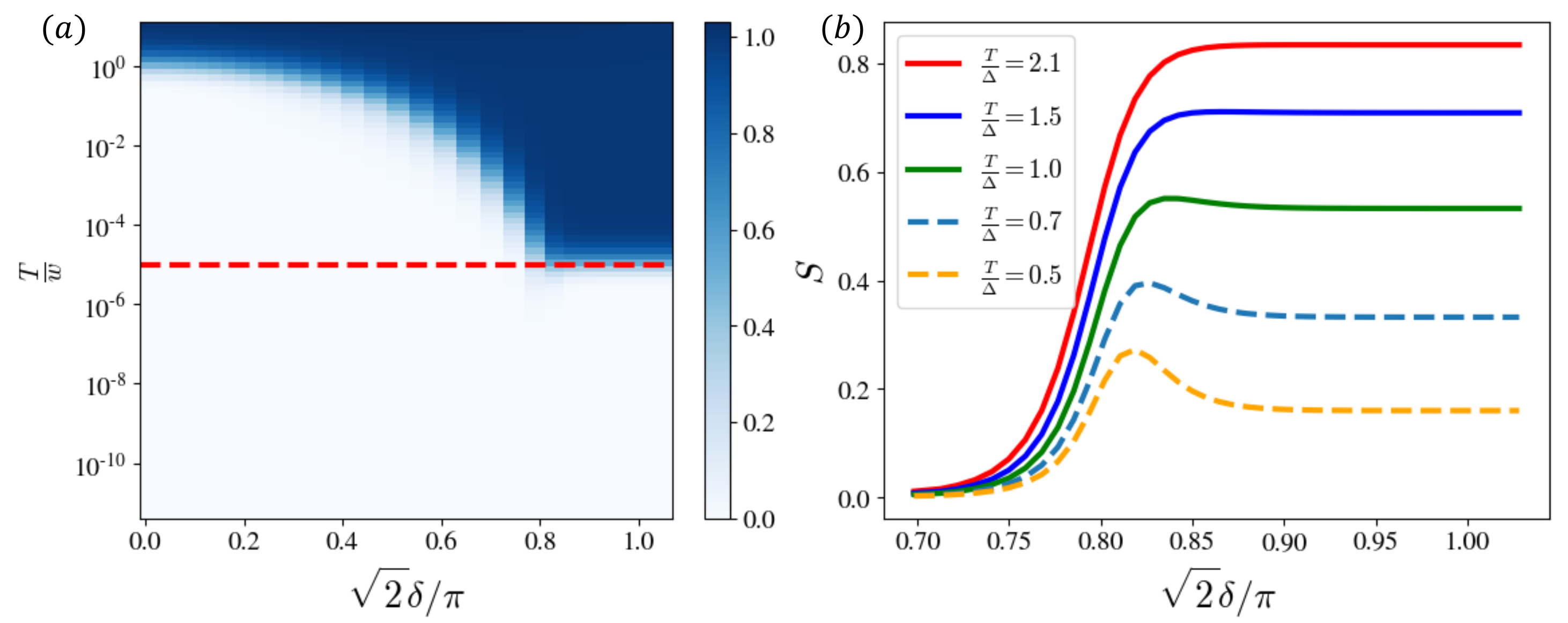}
    \caption{ Results for DD with finite asymmetry $\Delta = 10^{-5}w$. In panel (a), we show the phase diagram as a function of temperature $T/w$ and phase shift $\sqrt{2}\delta/\pi$. In panel (b), we show how $S$ depends on $\sqrt{2}\delta/\pi$ at various temperatures close to the asymmetry $\Delta$ (red cut in panel (a)).}
    \label{fig:3}
\end{figure}

We perform NRG calculations with $\Delta = 10^{-5}w$. In Fig.~\ref{fig:3} we show the entropy of the system.  All other parameters remain the same as for the $\Delta=0$ case in Fig.~\ref{fig:my_label1}(a). We see that no finite entropy phase survives below the temperature $T=\Delta$ (dashed horizontal line in Fig.~\ref{fig:3}(a)). Physically, below this temperature, the electron stays in the quantum dot with lower energy and is no longer affected by the QPC.

Interesting features emerge if we zoom in on the regime $T\sim\Delta\sim T^*$, where 3 different energy scales compete. In Fig.~\ref{fig:3}(b) we plot the entropy as a function of $\delta$ at various temperatures $T$ around $\Delta$. We observe a small peak developing as we  decrease temperature. The entropy then saturates to a finite value at large $\delta$.

The plateau developed at large $\delta$ can be explained simply by setting $w=0$ in the DD Hamiltonian, yielding an asymmetric two-level system.
However, the peak cannot be explained within a decoupled two-level system 
model.

In this section, we extend our  analysis to the case in which not only the scattering phase shifts but also the basis of  scattering states depends on $\hat{N}_A$. We will demonstrate that this model also undergoes a LT.

Specifically we consider the Hamiltonian $H=H_{DD} + H^{(f)}_{QPC}$, where
\bea
\label{non_symmetric_basis}
H^{(f)}_{QPC} &= -t'\sum_{i=-\infty}^{0}(f_i^{\dag}f_{i+1}+h.c.) -t'\sum_{i=1}^{\infty}(f_i^{\dag}f_{i+1}+h.c.) + t_{LR}(f_0^{\dag}f_1+f_1^{\dag}f_0)  \nonumber \\
 &+ V_0(f_{-1}^{\dag}f_{-1} + f_{0}^{\dag}f_{0} + f_{1}^{\dag}f_{1} + f_{2}^{\dag}f_{2}) 
+ V_L\hat{N}_A(f_{-1}^{\dag}f_{-1} + f_{0}^{\dag}f_{0}) + V_R\hat{N}_A(f_{1}^{\dag}f_{1} + f_{2}^{\dag}f_{2}), 
\eea
see Fig.~\ref{fig:tightbindingchain}(c).
This describes two semi-infinite chains connected by hopping amplitude $t_{LR}$, with a background scattering potential $V_0$, and a pair of $\hat{N}_A$-dependent potentials $V_L$ and $V_R$. If $V_L=V_R$, the scattering matrix is diagonal in the even/odd basis, $f_{e/o}\propto f_i \pm f_{1-i}$, regardless of the value of $\hat{N}_A$. In this case, only the scattering phase shift changes due to electron hopping in or out of QD $A$. On the contrary, if $V_L\neq V_R$, only when $\hat{N}_A=0$, $H^{(f)}_{QPC}$ is parity symmetric, but when $\hat{N}_A=1$, $H^{(f)}_{QPC}$ is not parity symmetric and the basis in which the scattering matrix is diagonal is not the even/odd basis. In this case, we cannot write down an effective Hamiltonian as simple as Eq.~(4) in the main text.\par

\begin{figure}
    \centering
    \includegraphics[width=.7\linewidth]{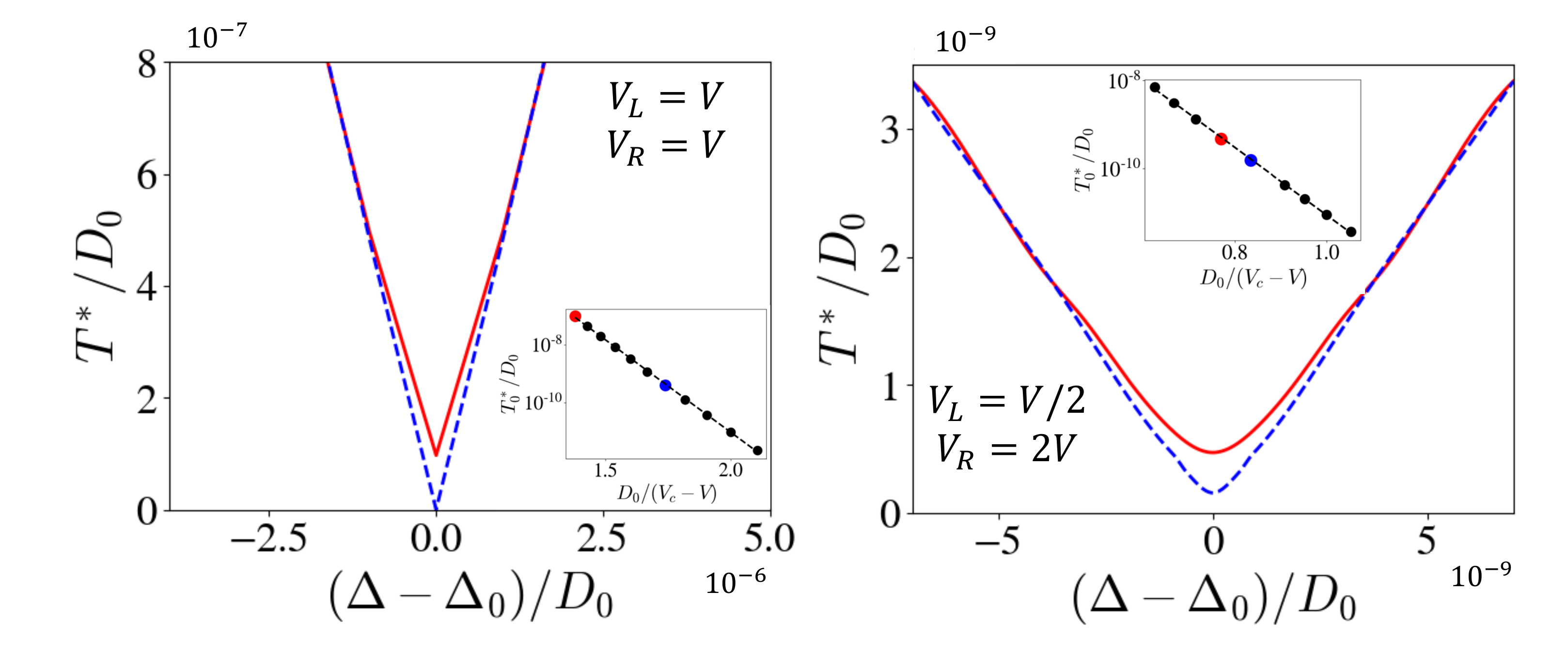}
    \caption{Crossover temperature as a function of DD asymmetry for the model in Eq.~(\ref{non_symmetric_basis}).}
    \label{fig:my_label11}
\end{figure}

We perform NRG calculations on $H$, with the following parameters: $t'=D_0=1.0$, $t_{LR}=0.2$, $V_0=-0.5$ and NRG discretization parameter $\Lambda=2.0$. We keep $400$ states at each iteration. Notice that the two QDs are not a priori symmetric: the asymmetry $\Delta$ needs to be fine-tuned such that $\langle{\hat{N}_A}\rangle=\frac{1}{2}$. In Fig.~\ref{fig:my_label11} , we show how the crossover temperature $T^{*}$ depends on $\Delta$, in both symmetric ($V_L=V_R=V$) and asymmetric ($V_L=V/2, V_R=2V$) case. We find that $T^*$ first decreases, and then increases, reaching a minimum at a value of $\Delta$ denoted as $\Delta_0$, which corresponds effectively to a symmetric DD.\par

Notice that when the scattering basis changes, the Anderson orthogonality catastrophe is not fully captured by the change of scattering phase shifts. Nevertheless, we  calculate the ground state overlap
\be
\langle GS(\hat{N}_A=0)|GS(\hat{N}_A=1)\rangle\sim L^{-x_b}
\label{eq:gsoverlap}
\ee
numerically, to extract the scaling dimension $x_b$ of the boundary condition changing operator. In Eq.~(\ref{eq:gsoverlap})  $|GS(\hat{N}_A=0(1)\rangle$ denotes the ground state of $H^{(f)}_{QPC}$ with $N_A=0(1)$. $L$ denotes the lattice size. At the critical value of $V$, $V_c$, we have $x_b(V_c)=1$ by definition. Based on the numerical value of $V_c$, we find the crossover temperature at the symmetric point, $T^{*}_0\equiv T^{*}(\Delta_0)$, decreases with $V$ exponentially, $T^{*}_0\sim exp(-const/(V_c-V))$, as shown in the insets. This confirms that $V_c$ is a quantum critical point.

\section{Effective theory}
In the following, we discuss various results associated with our effective theory,
\be
H_{{\rm{eff}}}(\phi) = w\sigma^x - \frac{v_F\delta}{\pi}\sigma^z\partial_x\phi(0) + \frac{v_F}{4\pi} \int dx (\partial_x\phi)^2.
\ee

{\bf{The QPC bath is ohmic:~}} 
Using imaginary time path integral, we can integrate out the boson field $\phi$. Then we obtain a new term in the effective action, given by
\be
\delta S \propto \int d\tau d\tau' \frac{\sigma^z(\tau)\sigma^z(\tau')}{|\tau - \tau'|^2},
\ee
where $\tau$ denotes the imaginary time, and the kernel $1/|\tau - \tau'|^2$ is set by the propagator of the operator coupled to $\sigma^z$, i.e., $\partial_x\phi(0)$. The same effective action describes the spin-boson model with Ohmic bath, see Ch.~3 of Ref.~\onlinecite{altland2010condensed}.
Similar arguments can be drawn for the RLM with a capacitively coupled QPC.

{\bf{Derivation of the RG equation for $\delta$:}}~Here, we apply the operator product expansion (OPE) method to derive the second equality in Eq.~(6) in the main text. In the following, we set $v_F=1$ for convenience.
First, applying the unitary transformation $U' = e^{i\sigma^z\frac{\delta}{\pi}\phi(0)}$ to the effective Hamiltonian $H_{{\rm{eff}}}(\phi)$, we have
\be
U H_{{\rm{eff}}}(\phi)U^{\dag} = w(\sigma^+e^{i\frac{2\delta}{\pi}\phi(0)} + \sigma^-e^{-i\frac{2\delta}{\pi}\phi(0)}) + \frac{1}{4\pi}\int dx (\partial_x\phi)^2.
\ee
From the OPE of two vertex operators, \cite{von1998bosonization}
\be
e^{i\lambda \phi(z)}e^{-i\lambda \phi(z')} = \frac{1}{(z - z')^{\lambda^2}} + \frac{\lambda i  \partial_{z'}\phi(z')}{(z - z')^{\lambda^2 - 1}} + \cdots,
\ee
we find that the OPE between $\sigma^{+}e^{i\frac{2\delta}{\pi}\phi(0)}$ and $\sigma^{-}e^{-i\frac{2\delta}{\pi}\phi(0)}$ gives us a term $\sigma^z\partial_x\phi(0)$ with OPE coefficient $\frac{2\delta}{\pi}$. Applying $U'$ again, this transforms back to the $\sigma^+e^{i\frac{2\delta}{\pi}\phi(0)}$ term. Following 
Ref.~\cite{cardy1996scaling}, we find that $\delta$ obeys
\be
\delta\to \delta - 2w^2\delta dl
\ee
under scaling transformation $a\to a(1+dl)$, where $a$ denotes the UV cutoff. This yields the desired result.

	\bibliography{QIbasedNonAbelian}